# Plant Defense Multigene Families: II Evolution of Coding Sequence and Differential Expression of PR10 Genes in *Pisum*.

Sandhya Tewari[1], Stuart M. Brown[2], Pat Kenyon, Margaret Balcerzak[3] and Brian Fristensky[*]

Department of Plant Science, University of Manitoba, Winnipeg, R3T 2N2, Canada
Current addresses: [1]Confederation of Indian Industry, 23 Institutional Area, Lodi Road, New Delhi - 110 003 INDIA
[2]Cell Biology Department, NYU Medical Center, 550 First Avenue, New York, NY 10016, U.S.A.
[3]Agriculture and Agrifoods Canada ECORC, 960 Carling Avenue, Ottawa, Ontario K1A 0C6, Canada

[*]Corresponding author:
Dr. Brian Fristensky, frist@cc.umanitoba.ca

While it is not possible to directly the observe evolution of multigene families, the best alternative is to compare orthologous family members among several closely-related species with varying degrees of reproductive isolation. Using RT-PCR we show that in pea (*Pisum sativum*) each member of the pathogenesis-related PR10 family has a distinct pattern of expression in response to the fungus *Fusarium solani*, and in treatment with salicylic acid, chitosan and abcisic acid. Sequencing reveals that PR10.1, PR10.2 and PR10.3 exist in *P. humile, P. elatius* and *P. fulvum*, except that no PR10.2 orthologue was identified in *P. elatius*. PR10.1, PR10.2 and PR10.3 appear to have diverged from a single gene in the common *Pisum* ancestor. For the recently diverged PR10.1 and PR10.2, the timing of fungal-induced expression differs greatly among species. For example, PR10.1 was strongly induced in *P. sativum* by *F. solani* within 8 hours postinoculation (h.p.i.), whereas little PR10.1 expression was seen in pea's closest relative, *P. humile*, and in the more distantly-related *P. elatius*. In *P. fulvum*, expression did not peak until 48 h.p.i. Expression of the more ancient PR10.4 and PR10.5 genes is more tightly conserved among *Pisum* species. These data indicate that expression, as well as sequence, can evolve rapidly. We hypothesize that changes in differential expression of multigene family members could provide a source of phenotypic diversity in populations, which may be of particular importance to plant/pathogen coevolution.

## INTRODUCTION

It is often taken for granted that many genes in plants are present in multigene families. Although it is difficult to be sure of the roles played by these families, there are several possibilities. For example, multiple copies of genes such as RUBISCO small subunit may facilitate the production of large quantities of gene product. In other cases, multigene families may allow the production of variants of a given protein, such as seed storage proteins. The observation that distinct copies of a gene may be differentiallly expressed with respect to other copies suggests that multigene families may be exploited by plants to facilitate more versatile regulatory regimes than are possible for single copy genes.

In a given species, temporal and developmental expression patterns can differ greatly between copies, implying that individual copies of a gene may be specialized for different functions. However, systematic study has never been done to determine whether copy-specific differential expression patterns are stable for each copy, or whether they diverge readily. In other words, does differential expression of members of a given multigene famliy represent a stable adaptation, or a transient evolutionary experiment?

Most defense-related proteins induced in plants in response to pathogens are encoded by multigene families, including phenylalanine ammonia-lyase [Cramer *et al.* 1989], chalcone synthase [Koes *et al.*, 1989), chalcone isomerase [Van Tunen *et. al.* 1988] hydroxyproline-rich glycoproteins [Corbin *et al.*, 1987], 4-coumarate CoA ligase [Douglas *et al.,*1987] β-1,3 glucanase (Ward *et al.*, 1991], PR1 [Rigden and Coutts, 1988), peroxidase [Harrison *et al.* 1995], and leucine aminopeptidase [Pautot et al., 1993]. In many cases, copies of a defense gene within a species tend to be more closely-related to each other (orthologous) than to copies from other species (paralogous). For example, phylogenetic analysis of thaumatin-like proteins (PR5) from oat and barley indicate that all four PR5 genes in oat cluster on one branch of the tree, while all barley sequences cluster together on a separate branch [Lin et al, 1996]. Clustering of gene copies within each species suggests that the extant copies of PR5 genes all descended recently from one PR5 gene present in the common ancestor of oat and barley. (The data do not distinguish between the same ancestral copy giving rise to PR5 genes in each species, versus different copies.)

Only a small number of studies have compared differential expression among individual members of defense multigene families in response to pathogens or elicitors [Chittoor et al., 1997, Junghans et al. 1993, Lin et al., 1996, Choi et al., 1994, Logemann et al., 1995, Båga et al. 1995, Danhash et al. 1993, Pérez-Garcia et al., 1995, Ward et al., 1991, Shufflebottom et al., 1993]. This is largely due to the difficulties involved in distinguishing transcripts from each copy of the gene. Generally, only a single multigene family in a single species was studied. However, Sun et al. [1997] have shown that five members of the polyubiquitin family exhibit both point mutations and differences in ubiquitin monomer repeats, as well as changes in copy-specific differential expression, between ecotypes of *Arabidopsis thaliana*.

These observations raise two questions: 1. Are orthologous copies of multigene family members conserved between closely-related species, or do gene copies turn over rapidly, such that there is no correspondence of gene copies from one species to the next? 2. Where orthologous copies are conserved, are differential expression patterns also conserved or do expression patterns for a gene change over short evolutionary times?

While it is not possible to observe the process of speciation directly, the best alternative is to compare species with varying degrees of reproductive isolation. *Pisum humile* is thought to be the wild pea from which *P. sativum* was domesticated [Waines, 1975]. Both spontaneous and artificial crosses among *P. sativum, P. humile* and *P. elatius* result in fertile offspring. However, crosses between *P. fulvum* and these three species result in either few offspring or offspring with greatly decreased fertilty [Ben-Ze'ev and Zohary, 1973] .

Using these pea species, we have previously shown [Tewari *et al*., 2003] that gene expression detected by a PR10.1/PR10.2 subfamily specific probe differs between *Pisum* species in response to the fungal pathogen *Fusarium solani*. However, the probe used in that analysis could not distinguish between PR10.1 and PR10.2 because of high sequence similarity between the genes. Here we report the cloning of members of PR10 genes from three wild *Pisum* species, for which most sequences are orthologous to PR10 genes from *P. sativum*. RT-PCR using gene-specific primers indicates that patterns of PR10 gene expression in response to *Fusarium solani* are divergent among *Pisum* species.

## MATERIALS AND METHODS
### Plant Material and treatments
Wild accessions of *Pisum* (*P. humile* 713, *P. elatius* 721 and *P. fulvum* 706) used in this study were obtained from N. O. Polans, Northern Illinois University, U.S.A. *P. sativum* c.v. Alaska was purchased from W. Atlee Burpee and Co., Warminister, PA. *F. solani* f. sp. *pisi* and *F. solani* f. sp. *phaseoli* were obtained from American Type Culture Collection (Accession numbers 38136 and 38135 respectively). Cultures were grown on potato dextrose agar (PDA) plates supplemented with a few milligrams of finely chopped pea leaf tissue.

All the *Pisum* and *Lathyrus* plants were grown in growth rooms in pots in 2:1:1 Soil:Sand:Peat mix under a day/ night cycle of 16/8 hours with temperatures of 22 /15 °C respectively. The average light intensity using 1/3 0-lux wide spectrum to 2/3 cool white was 340 μ e $m^{-2}$ $sec^{-1}$.

### DNA extraction from pea seedlings and young leaves
Pea hypocotyls and young leaves were frozen and lyophilized. Dry pea tissue was ground into powder in liquid $N_2$, then 1 ml of extraction buffer [100mM Tris-HCl (pH 8.0), 50mM EDTA, 500mM NaCl, 1.25% SDS] was added per 100 mg of tissue and incubated at 65°C for 20'. KOAc was added to a final concentration of 3M, the samples were kept on ice for 20' then centrifuged at 10X G for 15'. The supernatant was extracted twice with an equal volume of TE equilibrated phenol. DNA was precipitated with isopropanol and the pellet was dried and resuspended in TE at a concentration of 0.5 mg/ml.

### PCR conditions and cloning of PCR products.
PCR was performed in 25 μl using 1X buffer [50mM KCl, 10mMTris-HCl pH (8.0), 10mM NaCl, 0.01mM EDTA, 0.5mM DTT, 0.1% Triton X-100], 0.5 units Taq polymerase, 2mM MgCl, 40μM each of dNTP, 50 ng pea genomic DNA, 20 pmol of each primer, and 25 μl of mineral oil. A Techne PHC-2 unit was used with denaturation at 95° for 5', 35 cycles of 95° 1', 47° 2', 72° 2', and a final elongation at 72° for 10'. Products were electrophoresed in a 1.0% agarose gel and stained with EtBr. UV fluorescent bands were cut from the gel and DNA recovered Prep-A-Gene (Bio-Rad).

Isolated PCR products were TA-cloned by direct ligation into the pCRII vector (Invitrogen).

### DNA Sequencing
Single-pass sequencing was done for several clones per primer set, and a single clone was chosen for further sequencing. Sequencing of selected clones was done to at least 3-fold redundancy, either using the Vent DNA polymerase kit (Circumvent Sequencing kit, New England Biolabs) or by the DNA Sequencing Lab at the Plant Biotechnology Institute, National Research Council, Saskatoon, Canada.

### Computer analysis of DNA sequences
General sequence analysis tasks were carried out using the FSAP package [Fristensky et al. 1982], FASTA

programs [Pearson 1990], and XYLEM [Fristensky 1993].

Phylogenetic analysis was performed as follows: Protein coding regions (CDS) were extracted from GenBank [Burks et al., 1991] entries using the FEATURES program [Fristensky, 1993], and the corresponding amino acid sequences were aligned by PIMA [Smith and Smith, 1992], using maximal linkage and a cluster score cutoff of 25.0. Alignments of the original CDS sequences were performed using the PIMA protein alignment as input for MRTRANS [Pearson, 1990]. To produce the alignment in Figure 1, intron sequences were aligned separately using CLUSTALW 1.6 [Thompson et al., 1994], and then inserted into the alignment manually. 5' non-coding sequences were added and aligned manually. The DNA phylogeny was constructed aligned protein coding sequences (minus introns) using the maximum liklihood program fastDNAml 1.0.6 [Olsen et al., 1994] with 100 bootstrap replicates. Branch lengths were determined using the bootstrap consensus tree as input to fastDNAml. Trees were processed for figures using the TREETOOL tree editor [Maciukenas et al. 1994].

All programs were run from the Genetic Data Environment (GDE 2.3) [Smith et al. 1994].

**Pathogen inoculation and chemical treatments**
Immature pods (five pods per treatment) having no developed seed were harvested, slit longitudinally along the suture lines and placed with the freshly opened side up on a sterile petri-dish. Inoculation with $10^6$ macroconidia/ml of either *F. solani* f. sp. *pisi* or *F. solani* f. sp. *phaseoli* was done as described previously [Fristensky et al., 1985].

Chemical treatments were applied as for pathogen inoculations in a total volume of ten μl/pod half at the following concentrations: Chitosan, 1 mg/ml; ABA, 100 μM; and SA, 50 mM.

**RNA extraction**
Treated pod endocarp tissue was frozen in liquid $N_2$ and RNA extracted by the method of Verwoerd et. al. (1989) using the modifications described in [Tewari *et al.*, 2003].

**Reverse transcription**
Two μg of total RNA was incubated with 0.5 μg oligo $(dT)_{12-18}$ primer (Gibco BRL cat. # 18418-012) at 65 °C for 5 min. Reverse transcription was carried out in a 30 μl final volume at 50 °C for 30 min. in 50 mM Tris-HCl (pH 8.3), 75 mM KCl, 5 mM $MgCl_2$, 64 units of RNAsin (Gibco BRL), 12 units of AMV-RT (Promega), 1 mM each of dATP, dCTP, dGTP and dTTP.

**Internal control plasmids for RT-PCR**
pI49KSv was constructed by cloning the 585bp Sau3AI fragment from pUC18 into the BglI site within the PR10.1 cDNA in pI49KS [Tewari et al., 2003]. pI176KSiv was constructed by cloning the 585bp Sau3AI fragment from pUC18 into the BglI site within the PR10.1 cDNA in pI49KS [Tewari et al., 2003]. p49cKS contains the 868bp NsiI/XbaI coding sequence fragment from pCC2 [Chiang & Hadwiger, 1990], recloned into PstI/XbaI-digested BluescriptKSm13+. pABR17-10.1 was constructed as follows: the BamHI site from pBluescript KSm13+ was filled in using DNA polymerase Klenow fragment, and the resultant vector pMB5.2-2 was used to reclone the PR10.4 cDNA from pABR17 [Iturriaga et al., 1994], to give pABR17-10. The 141bp Sau3A1 fragment from pUC19 was next cloned into the BamHI site of the PR10.4 cDNA to give pABR17-10.1. pABR18-2.20 was constructed as follows: the PR10.5 cDNA was recloned from pABR18 [Iturriaga et al., 1994] into the KpnI site of pUC19 to give pABR18-2. Finally, the 245bp AluI fragment from pUC19 was cloned into the EcoRV site in the PR10.5 cDNA, to give pABR18-2.20. More details of constructs are found in Figure 3.

**DIG labelling of cDNA using PCR**
Ten μl of a 1:10 dilution of the cDNA synthesized using the method described above was used in the PCR reaction with specific primers (Table 1) for PR10.1 (oS49a+8 and oS49a-7), PR10.2 (oS49b+8 and oS49b-7), PR10.3 (oS49c+4 and oS49c-5), PR10.4 (oSABR17+4 and oSABR17-5) and PR10.5 (oSABR18+1 and oSABR18-5). PCR was carried out in a 25 μl total volume. Typically, 100 amole of internal control plasmid DNA was included, but the exact amount was adjusted empirically to avoid large discrepancies between mRNA-derived and control-derived band intensities.

PCR was carried out using the PCR DIG Labelling Mix from Boehringer Mannheim (Cat. # 1585 550) following manufacturer's instructions. The final concentration of the reaction mix was : 1X PCR buffer [10 mM Tris-HCl, 50 mM KCl (pH 8.3)], 1.5 mM $MgCl_2$, 200 μM dATP, dCTP, dGTP, 190 μM dTTP and 10 μM DIG-dUTP, 0.625 U Taq polymerase, 10 pmole of each primer. Wherever possible, master mixes were prepared to improve reproducibility. Fourteen cycles of PCR were carried out: denaturation at 94 °C, 1 min; annealing at 55 °C, 1 min; extension at 72 °C, 1.5 min.

**DIG Detection**
Five μl of the DIG labelled PCR product was electrophoresed on a 1.5% agarose gel and transferred to Hybond membrane (Amersham) following instructions from the manufacturer. The DNA was crosslinked to the membrane using the auto-crosslink mode of a Stratagene UV crosslinker. The blot was equilibrated in Buffer A [100 mM Tris-HCl (pH 7.0), 150 mM NaCl, and 0.3% Tween 20] for 1 minute and blocked in buffer B [1% (w/v) blocking reagent (Boehringer Mannheim cat. # 1096 176) in buffer A] for 30 min. on an orbital shaker. A 1:10,000 dilution of anti-DIG-AP (Cat. #1093274) conjugate in buffer B was prepared (final 37.5 U of anti-DIG-AP/ ml of buffer). This dilution was added and the membrane and incubated for 30 min., followed by two 15-min washes in buffer A. The membrane was then equilibrated in buffer C [100 mM Tris-HCl pH 9.5, 10 mM NaCl and 50 mM $MgCl_2$] for 2 min. A chemiluminescent substrate (1:100 dilution of a 25mM solution of CDP-Star, Cat. # 1685 627, in buffer C) was then added to the blot for 1 min. in a plastic bag. The solution was discarded and the blot exposed to X-ray film.

## RESULTS
*Pisum* **species contain conserved PR10 subfamilies**
To facilitate cloning of PR10 genes from wild pea species conserved and gene-specific PCR primers were created by inspection of a multiple alignment of all previously published PR10 sequences from pea, soybean, bean, potato and birch. Conserved primers were chosen from regions that exhibited minimal sequence divergence, and specific primers from more variable regions within the gene, as summarized in Table 1.

**Table 1.** Primers used for cloning and RT-PCR

| Gene | Primer name | Sequence | Posn |
|---|---|---|---|
| conserved | oC49+1 | 5'yawtityatcatgggtgt3' | -10 |
|  | oC49+3 | 5'cttactccaaaggttatt3' | 88 |
|  | oC49-5 | 5'aicagcatcacctttkgt3' | 483 |
|  | oC49-6 | 5'tttagttgtaatcaggat3' | 579 |
| Ypr10.1 | oS49a+4 | 5'ggtggtgctggaaccatcaaa3' | 143 |
|  | oS49a+8 | 5'ctagttacagatgctgataac3' | 67 |
|  | oS49a-5 | 5'atccccccttagctttgtcagt3' | 525 |
|  | oS49a-7 | 5'catccccccttagctttgtcag3' | 430 |
| Ypr10.2 | oS49b+4 | 5'ggaggtgctggaaccatcaag3' | 67 |
|  | oS49b+8 | 5'ctagttacagatgctgacact3' | 67 |
|  | oS49b-7 | 5'gcagcatcacctttttgtgtaa3' | 383 |
| Ypr10.3 | oS49c+4 | 5'tgttgaaggaaacggtggccc3' | 132 |
|  | oS49c-5 | 5'gatttcctcttcactaggaat3' | 395 |
| Ypr10.4 | oSABR17+4 | 5'ggtgatcaagaagaagcacaa3' | 99 |
|  | oSABR17-5 | 5'tttggcttttgtttcatcacg3' | 423 |
| Ypr10.5 | oSABR18+1 | 5'atgataccacctctaccgtcc3' | 23 |
|  | oSABR18-5 | 5'cttagctttgccttcctcaac3' | 423 |

Nomenclature: o = oligo; C = conserved; S = gene-specific; 49 refers to old gene designation "Drr49"; a = PR10.1-specific, b = PR10.2-specific, c= PR10.3-specific, + = forward, with respect to protein coding sequence, - = reverse; numbers following + or -are arbitrary. Ambiguities [Cornish-Bowden, 1985]: i = inosine, y = pyrimidine, w = A or T, K = G or T

Primer pairs for PR10.1, PR10.2, and PR10.3 were used to amplify PR10 coding sequences using genomic DNA from *P. elatius, P. humile* and *P. fulvum*. No PCR products were detected from any of the wild pea species when a PR10.2-specific primer pair oS49b+4 and oS49b-5 (5'ctcttcagtaggagcagcagc3'), not listed in Table 1) was used. Combinations of conserved and gene specific primers were needed to amplify putative PR10.2-specific PCR products, as indicated in Table 2. PCR products were cloned as described in Methods, and clones hybridizing with a PR10 probe were partially sequenced to identify PR10 genes. Nine clones were chosen for complete sequencing. Because the Ypr10.Ps.4 and Ypr10.Ps.5 sequences were not published until later in this work, homologues for these genes were not cloned.

An alignment of the nine PR10 sequences from wild peas and five previously-published PR10 sequences from *P. sativum* is shown in Figure 1. Sequences are grouped according to similarity. Despite the fact that all PR10.1 sequences amplified with the PR10.1 primers, and cluster together on the phylogenetic tree in Figure 2, it is difficult to conclude that they are strictly orthologous. Polymorphism between PR10.1 and PR10.2 sequences occurs at only 22 out of 384 positions between 142 and 525, the region over which all PR10 clones overlap. No base substitutions are seen exclusively in all PR10.1 sequences, or exclusively in PR10.2. Surprisingly, Ypr10.Pe.2, which had been amplified from *P. elatius* DNA using one PR10.2-specific primer (oS49b+4) and one conserved primer (oC49-6) is clearly most similar to other PR10.3 sequences than to PR10.1 or PR10.2. Thus, while no PR10.2 orthologue could be identified from *P. elatius*, two distinct PR10.3 sequences were amplified. Since Ypr10.Pe.2 and Ypr10.Pe.3 differ only at 6 positions, it may be that these two sequences are allelic, rather than distinct loci.

Amino acid polymorphism among PR10.1 and PR10.2 proteins was seen at only 9 out of 359 positions. The only amino acid insertion in *Pisum* PR10 proteins is an Alanine insertion corresponding to a GCT insertion in Ypr10.Ps.2 at positions 475-477.

```
                       -1        10        20        30        40        50        60        70        80        90
              oC49+1:yawtityatcatgggtgt   oSABR18+1:atgataccacctctaccgtcc                    oS49a+8:ctagttacagatgctgataac
                                                                                                          oC49+3:ctt
     ┌ Ypr10.Ps.1  CATTATCATCATGGGTGTTTTTAATGTTGAAGATGAAATCACTTCTGTTGTAGCACCTGCTATACTCTACAAAGCTCTAGTTACAGATGCTGATAACCTT
   1 │ Ypr10.Ph.1
     │ Ypr10.Pe.1
     └ Ypr10.Pf.1                                                                              oS49b+8:ctagttacagatgctgacact
     ┌ Ypr10.Ps.2  CAATATCATCATGGGTGTTTTTAATGTTGAAGATGAAATCACTTCTGTTGTAGCACCTGCTATACTCTACAAAGCTCTAGTTACAGATGCTGACACTCTT
   2 │ Ypr10.Ph.2
     └ Ypr10.Pf.2  TATTGTCATCATGGGTGTTTTTAATGTTGAAGATGAAATCACTTCTGTTGTAGCACCTGCTATACTCTACAAAGCTCTAGTTACAGATGCTGACACTCTT
     ┌ Ypr10.Ps.3  CATCATTATCATGGGTGTTTTCAATTTTGAGGAAGAAGCCACTTCCATTGTAGCTCCTGCTACACTTCTACAAAGCTCTGGTTACAGATGCTGACATTCTT
     │ Ypr10.Ph.3
   3 │ Ypr10.Pe.2
     │ Ypr10.Pe.3
     └ Ypr10.Pf.3
   4   Ypr10.Ps.4  TTTTTTTATCATGGGTGTCTTTGTTTTTGATGATGAATACGTTTCAACTGTTGCACCACCTAAACTCTACAAAGCTCTCGCAAAAGATGCTGACGAAATC
   5   Ypr10.Ps.5  ATCAATAATCATGGGTGTTTTCACATATGAGAATGATACCACCTCTACCGTCCCTCCTGCCAAGCTCTTCAAAGCTGTCGTGCATGACGCTGATCTCATC

                        100       110       120       130       140       150       160       170       180       190
              oC49+3:actccaaag                                 oS49a+4:ggtggtgctggaaccatcaaa
                                                               oS49b+4:ggaggtgctggaaccatcaag
              oSABR17+4:ggtgatcaaggaagcacaagg oS49c+4:tgttgaaggaaacggtggccc
                                                            TNTTGAAGGAAANGGTGGTGCTGGAACCATCAAGAAACTCACTTTCGTTGAAGgtcagtat-
     ┌ Ypr10.Ps.1  ACTCCAAAGGTTATTGATGCCATCAAAAGTATCGAAAT.G.........C.........A........................................
   1 │ Ypr10.Ph.1                                                         ............A....A...........................
     │ Ypr10.Pe.1                                                         .................A...........................
     └ Ypr10.Pf.1                                                         .................A..........................c
     ┌ Ypr10.Ps.2  ACTCCAAAGGTTATTGATGCCATCAAAAGTATCGAAAT.G..........C..A.......................................
   2 │ Ypr10.Ph.2                                                          ..A..........................................
     └ Ypr10.Pf.2  ACTCCAAAGGTTATTGATGCCATCAAAAGTATCGAAAT.G..........C..A..................A....A.............................c
     ┌ Ypr10.Ps.3  ACTCCAAAGGTTATTGATGCCATCAAAAGTATTGAAAT.........C.....CC.C.....................................t
     │ Ypr10.Ph.3                                         .G.......C.....CC.C...........................................t
   3 │ Ypr10.Pe.2                                         ..A.......C.....CC.C........A..............A...................t
     │ Ypr10.Pe.3                                         .G.......C.....CC.C........A....................................t
     └ Ypr10.Pf.3                                         .G.......C.....CC.C..G..........................................t
   4   Ypr10.Ps.4  GTCCCAAAGGTGATCAAGGAAGCACAAGGAGTCGAAAT.A.C........T..A...C..A...............G..AT.CA.TC......
   5   Ypr10.Ps.5  GTCCCAAAGGTTGTTGATTCAATCAAGACTGTTGAAATCC........T....C.A...C...TG........G......T......

                   200       210       220       230       240       250       260       270       280       290
                   --------------------------------------intron--------------------------------------------->
                   a-aat--atnc-t--t-tt-ac--ga-atat-c-t-t-anta-ta-tannatt-tt-a--a-t-tgnaat---t---t--tntgt-gcagATGGTGAAAC
     ┌ Ypr10.Ps.1  .....tt..a.a.ga......tt.......g...c.c.a.........aa..................c........a....................
   1 │ Ypr10.Ph.1  .....tt..a.a.g..g....tt.......g...c.c.a.........aa..................c........a....................
     │ Ypr10.Pe.1  .....tt..a.a.g..g....tt.......g...c.c.a.........aa..................c........a....................
     └ Ypr10.Pf.1  .....tt..a.a.g..g....tt.......g...c.c.a.........aa..................c........a...................C.
     ┌ Ypr10.Ps.2                                                                                         .........
   2 │ Ypr10.Ph.2  .....tt..a.a.g..g....tt.......g...c.c.a.........aa..................c........a....................
     └ Ypr10.Pf.2  .....tt..a.a.g..g....tt.......g...c.c.a.........aa..................c........a....................
     ┌ Ypr10.Ps.3  .g........t.........c.t....t......a.....t..c..g..tt...g..t.tg.t.g..a.....gaa.caa.tg.g...t.....C....
     │ Ypr10.Ph.3  .g...........t......g.......t........a.....t..c..g..tt...g..t.tg.c.g..a.....gaa.caa.gg.g...t.......
   3 │ Ypr10.Pe.2  .g...........t......g.......t........a.....t..c..g..tt...g..t.tg.c.g..a.....gaa.caa.gg.g...t.......
     │ Ypr10.Pe.3  .g...........t......g.......t........a.....t..c..g..tt...g..t.tg.c.g..a.....gaa.caa.gg.g...t.......
     └ Ypr10.Pf.3  .g.g.........t..........g.t.....t........a.....t..c..g..tt...g..t.tg.t.g..a.....gaa.caa.gg.g...t.......
   4   Ypr10.Ps.4                                                                                         ....AA....
   5   Ypr10.Ps.5                                                                                         GA..AC.G..

                          300       310       320       330       340       350       360       370       380       390
                   CAAGNATGTGTTGCACAAAGTGGAGTTAGTAGATGNTGCTAACTTGGCTTACAACTATAGCATAGTTGGNGGTGTTGGANTTCCAGACACAGTTGAGAAG
     ┌ Ypr10.Ps.1  ...AC..................T.............................T.................T......
   1 │ Ypr10.Ph.1  ...AC..................T.............................T.......T.C..............
     │ Ypr10.Pe.1  ...AC..................T.............................T.................T......
     └ Ypr10.Pf.1  ...C...................T.............................T.................T......
     ┌ Ypr10.Ps.2  ...AC..................T.............................T.................T......
   2 │ Ypr10.Ph.2  .....C.....C...........T.............................T.......C...T...........
     └ Ypr10.Pf.2  .....C.....C..........CT.............................T.................T.......
     ┌ Ypr10.Ps.3  ....T......A...........A..........G....AA.........C.........A.........C...G.....
     │ Ypr10.Ph.3  ....T......A...........A..........G...AAA.........C.........A.........C...G.....
   3 │ Ypr10.Pe.2  ....T......A...........A..........G....AA.........C.........A.........C...G.....
     │ Ypr10.Pe.3  ....T......A..........C.A..........G...AAA.........C.........A.........C...G.....
     └ Ypr10.Pf.3  ....T......A...........A..........G....AA.........C.........A.........C...G.....
   4   Ypr10.Ps.4  ...CT......C.A.......C.A..CGC...T....AA...A....T.G.........C...T.....A..A..ACCA..GC.A.AT...A.GTT.A......A
   5   Ypr10.Ps.5  .TT.T.C...............T...AGCCA.T......A....A..G..T.AA...T...T...C..T........A....C...TA.AT......T.........

                          400       410       420       430       440       450       460       470       480       490
                   ATCTCATTNGAGGCTAAACTGTCTGCAGGACCAAATGGAGGATCCATTGCAAAGCTGAGTGTGAAATATTACACAAAAGGTGAT---GCTGCTCCTANTG
     ┌ Ypr10.Ps.1  .....C.........................................................................C.T............C..
   1 │ Ypr10.Ph.1  ..................................................................................................C..
     │ Ypr10.Pe.1  .....T.......................................................A.........C.T...............C..
     └ Ypr10.Pf.1  .....C..........................................................................................C..
     ┌ Ypr10.Ps.2  .....T..........................................................................................GCT....C..
   2 │ Ypr10.Ph.2  ................................................................................................C..
     └ Ypr10.Pf.2  .....T.........................................................................................C..
     ┌ Ypr10.Ps.3  .....G..T.....T........................................T..C.................AT.....G.
     │ Ypr10.Ph.3  .....T..T.....T........................................T..C.................AT.....G.
   3 │ Ypr10.Pe.2  .....G..C.....T........................................................AT.....G.
     │ Ypr10.Pe.3  .....G..C...............................................................AT.....G.
     └ Ypr10.Pf.3  .....T..T............A.G...G...........................C....T..C................AT.....G.
   4   Ypr10.Ps.4  G.TG....C....A..TTA.T.TG..T..TT.TG.C..T........C.TT....A.ATC.....C...C............A....TATC.
   5   Ypr10.Ps.5  ..A.....T......C....T..A...T..........TG...GT...A...T....T...C.T........A......AAG....T..

                                                                 tgktttccacta  cgacia:oC49-5
                                                                       cgacgacgaggatgac:oS49b-5
                                                              aatgtgttttccactacgacg:oS49b-7
                                          cctagitaicgtttcgactcacamtttat:oC49-9              taaggatcac:oS49c-5
```

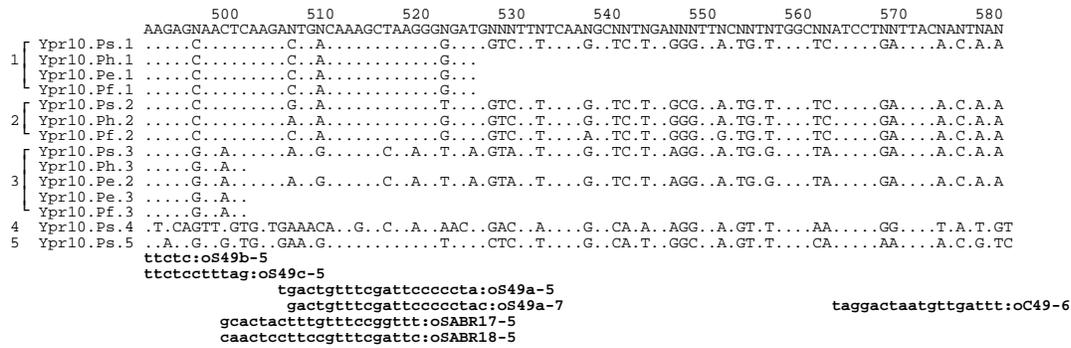

**Figure 1**. Alignment of *Pisum* PR10 coding regions.

Sequences are grouped top to bottom into subfamiies 1, 2, 3, 4 and 5. The alignment is written with reference to a consensus sequence, which appears at top. Periods (.) indicate positions that agree with the consensus, while letters indicate nucleotides that differ. Gaps are represented by dashes (-) and blanks represent positions for which no sequence information was available (eg. Ypr10.Ps.2, Ypr10.Ps.4 and Ypr10.Ps.5 are cDNAs, so introns are absent). Exons are in capitals and introns in lowercase. Forward primers are written 5' to 3' above the sequence and reverse primers are written 3' to 5' below the sequence.

Table 2. Cloned PR10 genes from *Pisum* sp.

| Gene | Species | Genbank ACCESSION | previous designation | Refrence[1] or primers used[2] |
|---|---|---|---|---|
| Ypr10.PS.1 | *P. sativum* | U31669 | Drr49a,pI49 | Culley *et al.*, 1995 |
| Ypr10.PH.1 | *P. humile* | U65419 | - | oS49a+4, oS49a-5 |
| Ypr10.PE.1 | *P. elatius* | U57064 | - | oS49a+4, oS49a-5 |
| Ypr10.PF.1 | *P. fulvum* | U65424 | - | oS49a+4, oS49a-5 |
| Ypr10.PS.2 | *P. sativum* | M81249 | Drr49b, pI176 | Fristensky *et al.*, 1988 |
| Ypr10.PH.2 | *P. humile* | U65420 | - | oS49b+4, oC49-6 |
| Ypr10.PF.2 | *P. fulvum* | U65425 | - | oC49+1, oC49-6 |
| Ypr10.PS.3 | *P. sativum* | J03680 | Drrg49-c | Chiang and Hadwiger, 1990 |
| Ypr10.PH.3 | *P. humile* | U65421 | - | oS49c+4, oS49c-5 |
| Ypr10.PE.2 | *P. elatius* | U65422 | - | oS49b+4, oC49-6 |
| Ypr10.PE.3 | *P. elatius* | U65423 | - | oS49c+4, oS49c-5 |
| Ypr10.PF.3 | *P. fulvum* | U65426 | - | oS49c+4, oS49c-5 |
| Ypr10.PS.4 | *P. sativum* | Z15128 | ABR17 | Iturriaga *et al.*, 1994 |
| Ypr10.PS.5 | *P. sativum* | Z15127 | ABR18 | Iturriaga *et al.*, 1994 |

[1] Citation for previously-cloned gene

[2] Primers used to amplify sequences used in this study

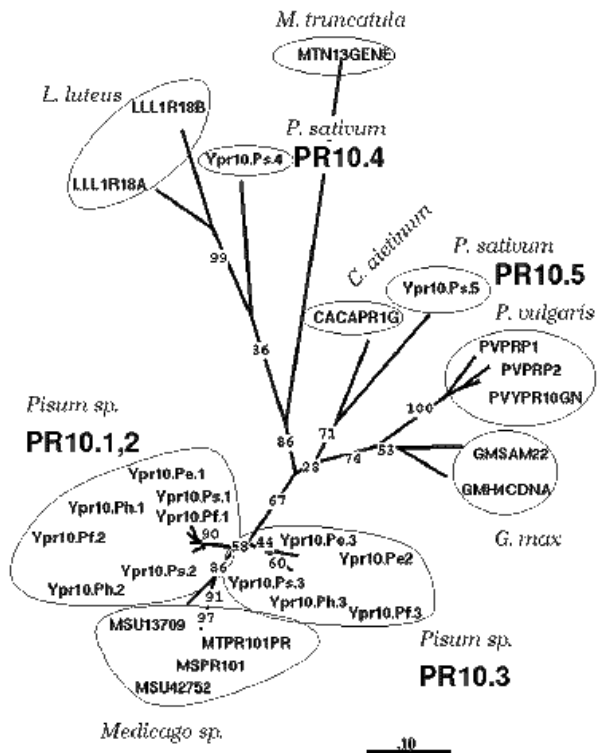

**Figure 2.** Phylogenetic relationships of legume PR10 genes by the method of maximum liklihood.

The tree topology represents the consensus of 100 bootstrapped replicates. Branch lengths were calculated using the entire alignment and the consensus tree as input. The percentage of replicates in which a group of sequences clustered together is indicated along each branch axis. Sequences are represented by GenBank LOCUS names, or by designations from Table 2.

A maximum liklihood tree (Figure 2) was constructed using protein coding sequences (ie. minus introns and flanking regions) aligned in Figure 1. Essentially the same topology was also obtained using either parsimony as implemented in DNAPARS or the distance method in FITCH [Felsenstein, 1985].

PR10 subfamilies in legumes tend to cluster within species. The clustering of PR10.1 and PR10.2 genes separately from PR10.3 suggests that both PR10.1 and PR10.3 were present in the common ancestor of all four *Pisum* species. Since all species have a PR10.1 gene, and all but *P. elatius* have a PR10.2 gene, it is also likely that a gene duplication event created the PR10.1,PR10.2 class of genes, prior to the divergence of these species. In this model, either PR10.2 was lost from *P. elatius* or the priming sites for this gene diverged sufficiently to prevent amplification with the conserved and gene-specific primer combinations tested.

Interestingly, four of the five PR10 genes from *Medicago* species cluster on a distinct clade roughly equidistant between *Pisum* PR10.1/PR10.2 and PR10.3. Also, in 86 out of 100 bootstrap replicates, MTN13GENE and Ypr10.Ps.4 cluster together. These data suggest that prior to the divergence of *Pisum* and *Medicago*, two ancestral genes were present, one of which gave rise to the PR10.1, PR10.2 and PR10.3 orthologues, and the other to PR10.4 and PR10.5 orthologues. The placement of PR10.4 and PR10.5 on separate clades is most consistent with the model that these genes represent discrete, paralogous copies of PR10 that diverged early in the evolution of legumes. Interestingly, there are two PR10.4-like genes in *M.Luteus*, while *P. vulgaris* and *G. max* have at least three and two copies, respectively, of PR10.5. It must be noted that these copies represent only published sequences. Other unsequenced copies may also exist.

**Specificity and linearity of RT-PCR assay**

Plasmid constructs were designed to serve as internal PCR standards. Figure 3A lists plasmids containing *P. sativum* PR10 genes, and constructs derived from these plasmids, containing inserts between the priming sites. When added to RT-PCR reactions, plasmid sequences should coamplify with the mRNA-derived PCR product. The presence of inserts within the amplified region results in plasmid-derived PCR products that have a higher molecular weight than the mRNA-derived product. Since no cDNA was available for PR10.3, the genomic clone itself, containing an 84bp intron, was used as a standard. Each internal standard generates a PCR product distinct from those for other genes. If a fixed molar quantity of standard is added to each RT-PCR reaction, the standard can serve as a control both for specificity and uniformity of amplification, from reaction to reaction.

The specificity of each primer pair was tested with each of the five internal standards. All the primer pairs detected only the respective sequences for which they were designed at low [100 a mole (1 amole= $10^{-15}$ moles)] template concentrations (data not shown).

It was important to test if the ratio of signal intensities of the detected bands represent the ratio of RNA amounts present in the beginning, since at higher number of cycles, transcripts which are present in low abundance are over-represented, while those present in

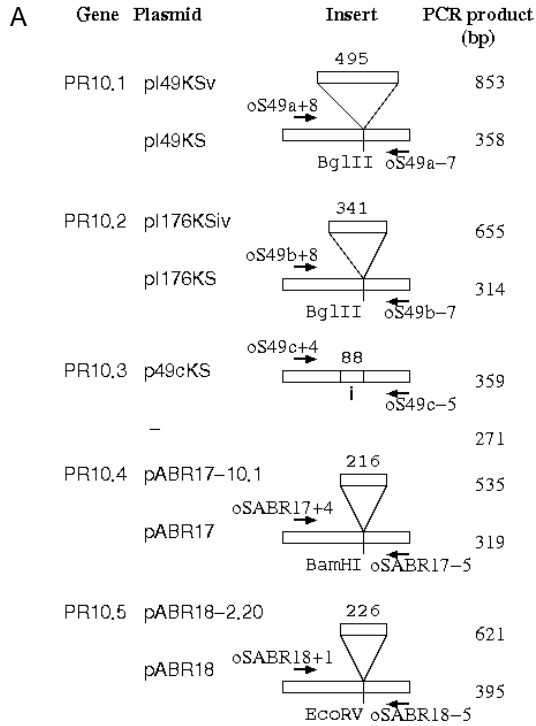

**Figure 3.** Controls for specificity and linearity of amplification.

**A.** Five control plasmids were constructed by cloning pUC19-derived inserts into PR10 coding regions, within the region to be amplified by the primers indicated in the figure. For PR10.3, p49cKS was constructed, containing an 868bp NsiI/XbaI fragment from the PR10.3 intron (**i**) -containing genomic clone. Specific details of plasmid construction can be found in the Methods section. Sizes of amplification products from the original plasmids (equivalent to the mRNA-derived product), and products derived from control plasmids, are listed at right.

**B.** Linearity of signal as a function of input DNA, after 14 cycles of PCR. Three-fold dilutions of a mixture of plasmid paris for each gene were amplified using gene-specifiec primer pairs indicated in A. Autoradiographic signal, as measured by densitometry, is plotted versus the input amount of DNA in attomoles. For each set of bands, the plasmid from which it was amplified is indicated at left. (▼, top bands) cDNA with insert; (△, bottom bands) cDNA without insert.

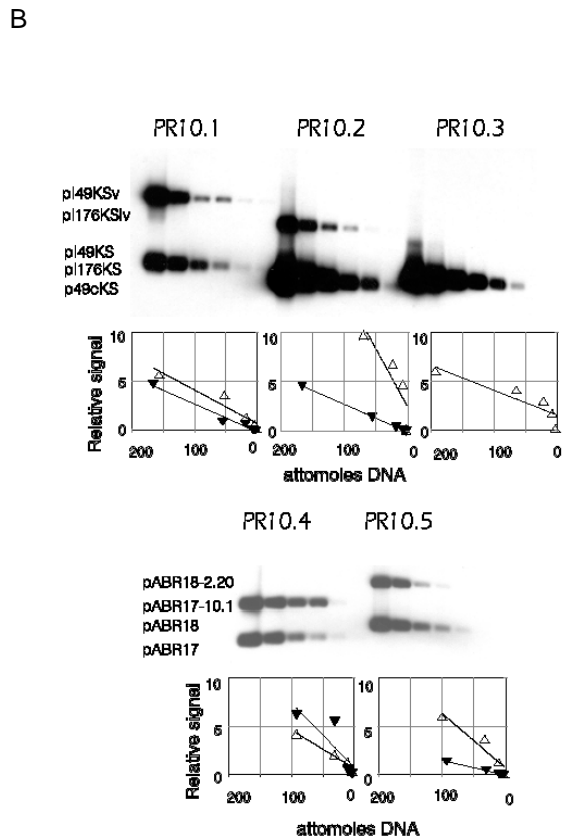

high levels will reach a plateau and hence be relatively under-represented. When all other reagents are in molar excess over PCR product, it is possible to obtain a linear relationship between template input and the output signal by limiting the amount of template and the number of PCR cycles. A dilution series of the cloned DNA plasmids, ranging from 1000 amoles to 2 amoles, was made. This series was subjected to 10, 14 and 17 cycles of PCR and a standard curve constructed (data not shown). Fourteen cycles of PCR was found to be sufficient in maintaining the range of assay linear without compromising the sensitivity of detection.

The specificity and linearity of this assay is illustrated in Figure 3B. For each gene, an approximately equimolar mixture of each cDNA and the cDNA with insert, as listed in Figure 3A, were amplified using 14 cycles of PCR. For example, in the gel at top, the leftmost set of lanes contains 6 three-fold dilutions of a mixture of pI49KS and pI49KSv, representing the PR10.1 gene. The bands in these lanes were amplified using the PR10.1-specific primer pair oS49a+8 and oS49a-7 (Figure 3A). For each of the primer pairs listed in Figure 3A, gene-specific amplification of PCR product is shown in Figure 3B. These results demonstrate that amplification of gene-specific PCR product is roughly linear over at least two orders of magnitude. At initial DNA concentrations below about 10 attomoles, signal was not always seen. Similarly, above about 200 attomoles, the slope of the dilution curve drops off. This is probably in part due to saturation of the film with high signal. For this reason, only the datapoints in the 0 to 200 attomole range are plotted.

While the relationship between input DNA and autoradiographic signal is linear between 10 - 200 attomoles of template, the slope of the line determines the amount of increase in signal per attomole of DNA added. For example, with pI176KS, the dilution from 70 amole down to 8 amole results in a 2-fold decrease in band density. At the other extreme, with pABR18-2.20, dilution from 93 to 10 attomoles resulted in a 23-fold decrease in band intensity. These examples demonstrate that differences in autoradiographic signals underestimate the underlying differences in target sequences being quantitated. That is, small differences in target DNA between treatments will not result in discernable differneces in autoradiographic signal. Large between-treatment differences in target sequence will be required to give obvious differences in signal, using this assay.

Prior to using RNA samples for RT-PCR, RNA was quantitated by either absorbance at 260nm or flourimetry. Based on these readings, equal amounts of RNA from each treatment were electrophoresed and compared for equal intensity by EtBr staining. Dilutions of samples were made to correct for differences in intensity, and the samples were checked again by electrophoresis. This process was repeated, as many as four times, until all samples showed roughly equal intensity in EtBr staining (data not shown).

Finally, to ensure that the RT-PCR assay was detecting products amplified from RNA rather than co-purifying genomic DNA, total RNA samples that had not been subjected to cDNA synthesis were added to PCR reactions. After 14 cycles of PCR with either PR10.4 or PR10.5-specific primers, no labeled PCR products were detectible (data not shown).

**Time course of PR10 transcript accumulation in *P. sativum***

Distinct differences in expression of PR10 genes were apparent in three independent timecourse experiments with *F. solani* f. sp. *phaseoli* (resistance response) and *F. solani* f. sp. *pisi* (susceptible). PR10.1 PCR products could be detected in autoradiograms as early as 2 hours after inoculation (Figure 4). Transcript levels increased sharply within 4 hours, reaching a peak by 8-12 hours. PR10.2 transcript was not detectable until 8 hours, and did not reach peak levels until 32 hpi. In all experiments, PR10.2 signal was substantially weaker than PR10.1. (In this case, signal can be compared between the two genes because they were always loaded on the same gel.) These results are consistent with previously-published studies, in which PR10 mRNA accumulation, as detected by hybridization of cDNA to total RNA, peaked by 8 hpi after inoculation with either fungus [Fristensky et al., 1985]. Since the PR10.1 and PR10.2 probes used in that study cross hybridized, it is likely that the PR10 mRNA induction detected in the first 8 hpi was largely due to PR10.1 transcript.

PR10.3 transcripts were never detected in any of these experiments, even though PR10.3 internal controls were detectible in all experiments at levels comparable to other controls. In Tewari et al., 2003, we demonstrated that autoradiographic signals detected by a PR10.3-specific cDNA probe were routinely much weaker than bands detected by PR10.1/PR10.2-specific probes, even when longer exposure times were used. In RNA samples from *P. sativum* treated with *F. solani* f. sp. *phaseoli* for 8 hr., 25 cycles of PCR were required to detect the 271bp fragment derived from the PR10.3 mRNA [data not shown]. In controls from which reverse transcriptase was omitted from the cDNA synthesis reaction, the 271bp band was not detected. In both experiments, a 359bp band, comigrating with the intron-containing genomic fragment from p49cKS, was detected after 20 cycles of PCR [data not shown]. Presumably this band, which was not detected after 14 cycles, was derived from trace amounts of genomic DNA in the RNA preparations. We conclude that the PR10.3 transcript, while present in *Fusarium*-inoculated pod tissue, is at too low a concentration to be detected in the linear range of the assay used here.

PR10.4 RT-PCR products were minimal or undetectable at 2 hpi, and generally weak until 8 hpi. Peak expression was typically seen by 12 hpi. In contrast to the other genes, PR10.5 accumulation was strong by 2 hpi, and peaked within 8-12 hpi. PR10.5 expression also declined significantly by 48 hpi.

While expression patterns were in general similar for a given gene with either fungus, some race-specific differences were seen. The most striking difference was seen with PR10.2, whose expression was typically weaker with the compatible *F. solani* f. sp. *pisi* than in an incompatible interaction with *F. solani* f. sp. *phaseoli*. With *F. solani* f. sp. *pisi*, PR10.2 expression dropped almost to baseline levels at 24 hpi but recovered to peak levels by 32 hpi. Previous studies using PR10.1 or PR10.2 cDNA probes also detected a comparable drop in PR10 mRNA levels at 24 hpi, followed by an increase in expression by 48 hpi. [Fristensky et al., 1985]. Since the PR10.1 and PR10.2 probes cross hybridized, the results from that paper must be interpreted as the sum of the mRNA accumulation for both genes, such that a substantial drop in mRNA level for PR10.2 would have been detected by either probe. PR10.4 and PR10.5 also exhibit modest decreases in mRNA levels at 24 hpi. with *F. solani* f. sp. *pisi*. However, peak expression resumes by 32 hpi for PR10.4, while PR10.5 exhibits no return to peak levels at later hours. Finally, PR10.1 transcripts accumulate more rapidly with the incompatible *F. solani* f. sp. *phaseoli*, with stronger expression at 2, 4, and 8 hpi, as compared to interactions with *F. solani* f. sp. *pisi*.

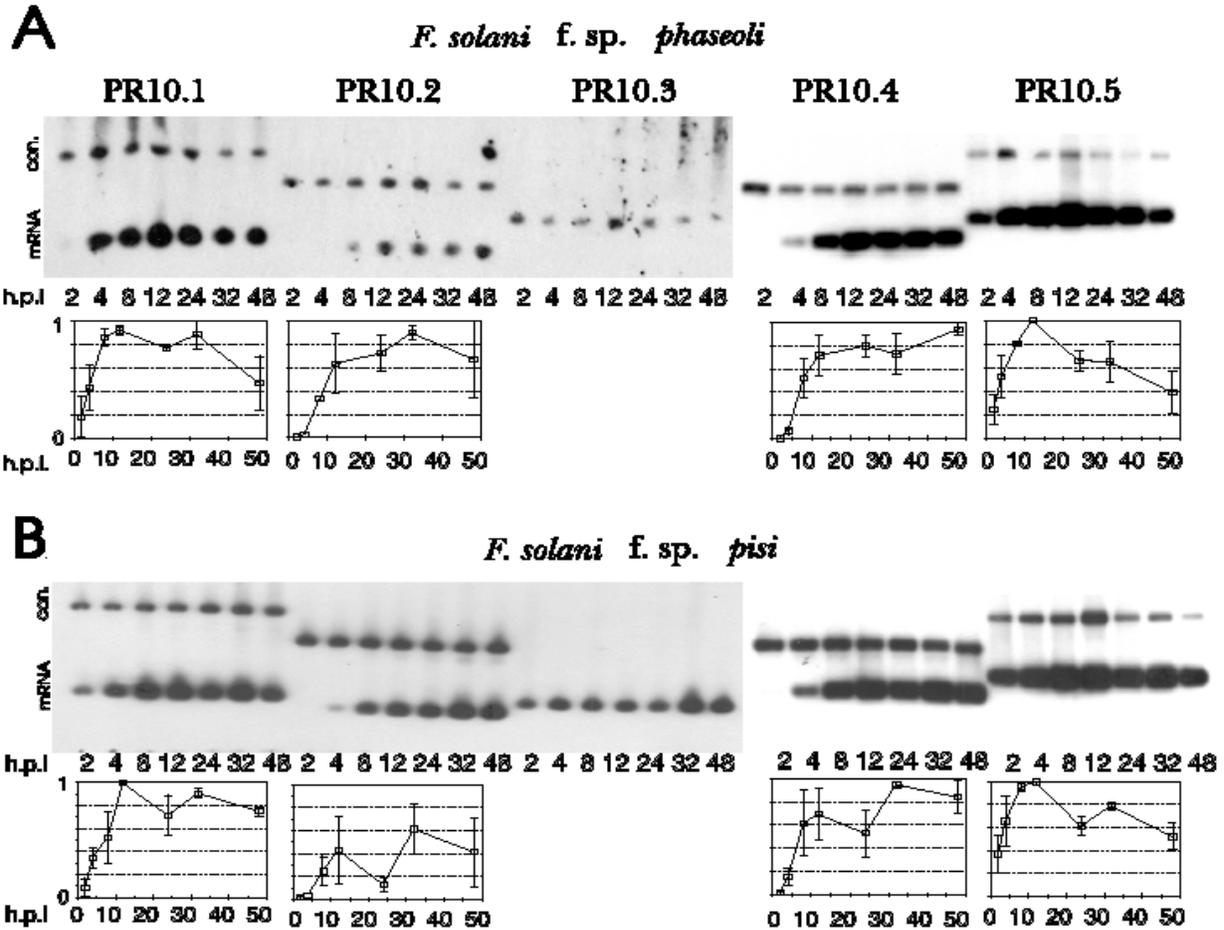

**Figure 4.** Timecourse of accumulation of specific PR10 mRNAs in *P. sativum*.

**(A)** In response to *Fusarium solani* f. sp. *phaseoli*
**(B)** In response to *Fusarium solani* f. sp. *pisi*.
cDNA synthesized from reverse-transcription of RNA pod tissue treated with fungus for the indicated time was amplified by RT-PCR. and products were electrophoresed, transferred to membranes, and DIG-labeled products detected using a chemiluminescent substrate. For each gene, the PCR product amplified from the internal standard plasmid migrates at a higher molecular weight class than the mRNA-derived product. Histograms represent relative autoradiographic signal, as measured by densitometry, averaged over at least three experiments. Vertical lines indicate the standard error of the mean.

### Differential transcript accumulation in *P. sativum* in response to chemical treatments

To determine whether PR10 expression is inducible by treatments other than pathogen challenge, pea pods were treated with salicylic acid, abscisic acid, chitosan, or water as a control. Only PR10.4 and PR10.5 showed detectible expression in water treated pods. With salicylic acid, PR10.4 and PR10.5 also gave some signal, but levels were close to those in the water control. Chitosan treatment resulted in induction of PR10.1, PR10.4 and PR10.5 within 8 hpi. PR10.4 transcript was detectable at both 8 and 48 hours, whereas PR10.5 expression declined after 8 hpi. PR10.2 transcripts were barely detected following treatment with this elicitor, and PR10.3 was not detected.

Following abscisic acid treatment, both PR10.4 and PR10.5 were induced, with transcript levels increasing between 8 and 48 hours (Figure 5). Only faint signal was detectible for PR10.1 transcripts in ABA-treated tissues. No signal was observed for PR10.2, PR10.3.

### Differential expression of PR10 genes in *Pisum* species

Having demonstrated that at least three PR10 genes are conserved in wild pea species, we were interested in knowing if the differential expression patterns for PR10

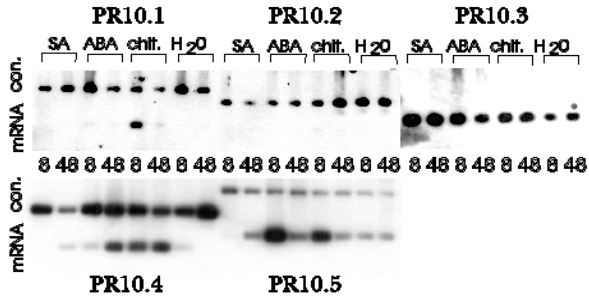

**Figure 5**. Accumulation of specific PR10 mRNAs in *P.sativum* in response to chemical treatments.

RNA isolated from pod tissue treated with 50 mM salicylic acid (SA), 100 μM abscisic acid (ABA), 1 mg/ml chitosan or water for indicated times was assayed as described in Figure 4.

genes were also conserved across the genus. RNA was isolated from *P. sativum, P. humile, P. elatius* and *P. fulvum* pod tissue treated with *F. solani* f. sp. *phaseoli* or *F. solani* f. sp. *F. solani* f. sp. *pisi*. RT-PCR was performed using specific primers for PR10.1-5, as described in Methods. The PR10.1-3 signals are directly comparable since they were loaded together on a single gel in all experiments as were PR10.4 and PR10.5.

PR10.1
PR10.1 transcript accumulated in both *P. sativum* and *P. fulvum* in response to challenge with either *F. solani* f. sp. *pisi* or *F. solani* f. sp. *phaseoli*, although the kinetics of accumulation were different in each species (Figure 6). Further, *P. fulvum* showed differences in relative abundance of PR10.1 mRNA upon challenge with the two pathogens. *P. humile* accumulated this transcript in response to inoculation with *F. solani* f. sp. *pisi* only and not with f. sp. *phaseoli*. One puzzling observation is that PR10.1 signal was not detectible in *P. humile/F.solani* f.sp *pisi* interactions in some experiments (ex. Figure 6B, autoradiogram) while in other experiments, signal was seen at 48 h.p.i (Figure 6B, histogram). PR10.1 mRNA was not detectable with either pathogen in *P. elatius*. Although the pattern of accumulation was similar in response to both pathogens, *P. fulvum* accumulated much higher levels of this transcript with *F. solani* f. sp. *pisi*.

PR10.2
PR10.2 was expressed in *P. sativum* and its closest relative *P. humile* upon challenge with either pathogen (Figure 6), although expression in response to *F. solani* f. sp. *phaseoli* was very weak. In *P. sativum*, the 8 h.p.i./48 h.p.i. ratio varied substantially between experiments, as indicated by comparing the autoradiogram with the histogram. Some mRNA was also detectable in *P. fulvum* inoculated with *F. solani* sp. *phaseoli*. Very little expression was seen in *P. elatius*.

PR10.3
The transcript for this gene was not detected in any of the host species upon infection with either *F. solani* f. sp. *phaseoli* or *F. solani* f. sp. *pisi* (Figure 6) although the internal control amplifies using the PR10.3 specific primer.

PR10.4
All host species accumulated PR10.4 mRNA when inoculated with either *F. solani* f. sp. *phaseoli* or *F. solani* f. sp. *pisi* (Figure 6). In most species, strong signal was seen at both 8 and 48 hours. *P. elatius* showed a significant induction of PR10.4 at 48 h.p.i with both fungi, and *P. sativum* and *P. fulvum* only showed a significantly stronger 48 h.p.i induction with *F. solani* f. sp. *phaseoli*.

PR10.5
High levels of PR10.5 mRNA were detected in the *F. solani* f. sp. *phaseoli* -treated pod tissue of all the hosts with both fungi (Figure 6). Except in *P. elatius*, where similar levels of mRNA were present at both time points tested (8 and 48 hours), in all other species, both fungi usually induced strong expression by 8 h.p.i, with decreased expression at 48 h.p.i. One execption was *P. fulvum*, in which *F. solani* f. sp. *pisi* induced roughly the same mRNA levels at 8 and 48 h.p.i.

The timecouse data from Figure 4 can be used as a check of the results for *P. sativum* in Figure 6. For a given gene/pathogen combination, the ratio of the 8 and 48 hour timepoints generally agree, within the range of the standard error of the mean, between Figure 4 and Figure 6. For example, the mean transcript levels for PR10.4 at 8hpi is less than the mean transcript levels at 48 hpi. for both fungi, in Figures 4A and 6A. Similarly, PR10.5 transcript levels are greater at 8 hpi than at 48 hpi, for both fungi (Figure 4A, Figure 6A). The data for PR10.1 and PR10.2 with *F. solani* f. sp. *phaseoli* are also in agreement between Figure 4A and Figure 6A. However, the data for PR10.1 and PR10.2 in plants treated with *F. solani* f. sp. *pisi* are not as consistent as the rest of the data. Although the 48 hpi time point for PR10.1 in Figure 4B timecourse has a negligible standard error, the 8 hpi time point has substantial variation. Therefore, this 8 hour timepoint can not be used to corroborate the results in Fig 6B. Similarly, both the 8 and 48 hour timepoints for PR10.2 in Figure 4A have overlapping standard errors. Therefore, although both PR10.1 and PR10.2 in *P. sativum* show greater mean levels of mRNA at 8 hpi vs. 48 hpi, these differences may not be significant. (Note that it would

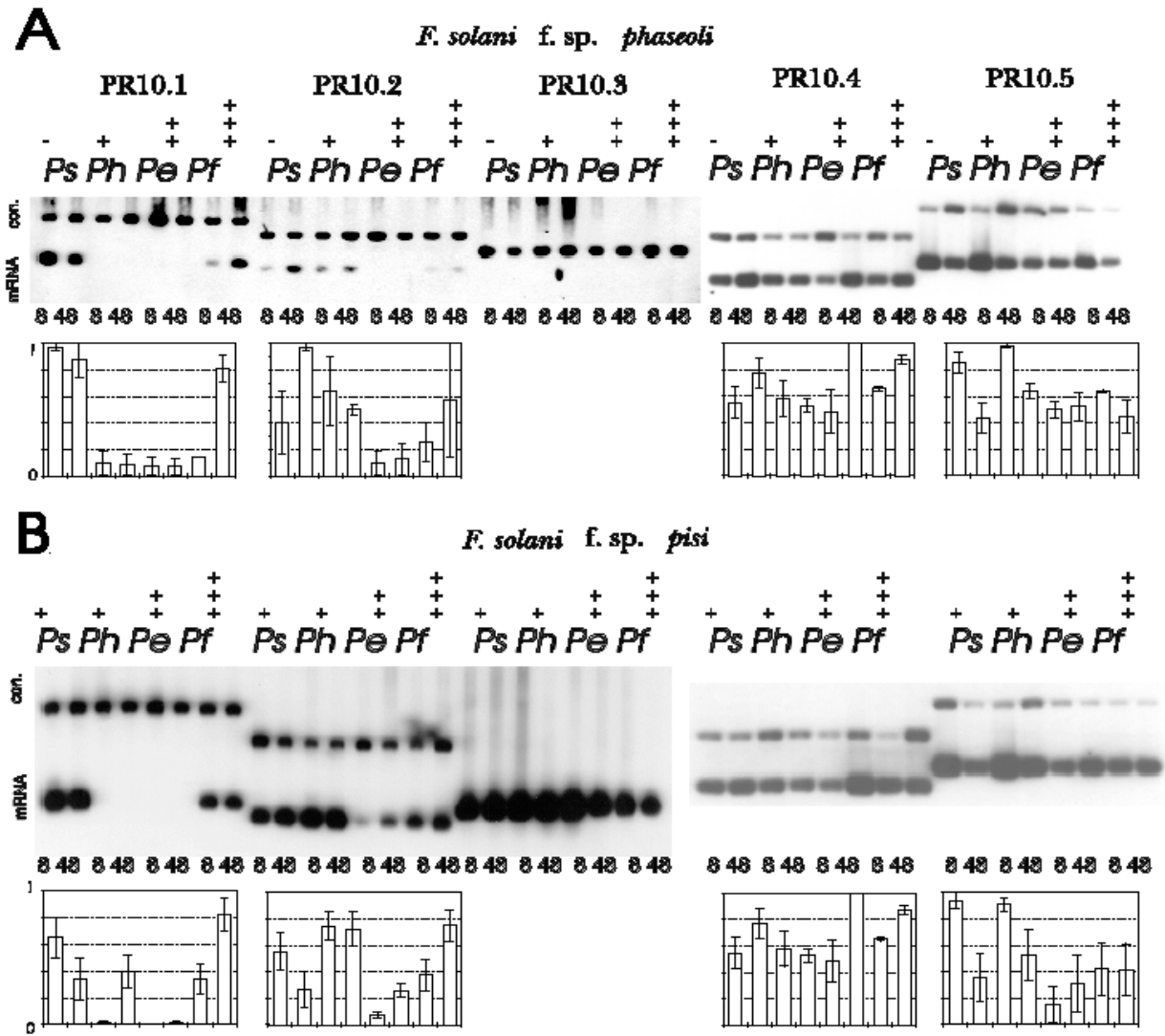

**Figure 6.** Differential accumulation of specific PR10 mRNAs in *Pisum* species.

**(A)** In response to *Fusarium solani* f. sp. *phaseoli*
**(B)** In response to *Fusarium solani* f. sp. *pisi*.
RNA isolated from fungus-treated pod tissue from *P. sativum, P. humile, P. elatius* and *P. fulvum* was assayed as described in Figure 4. Extent of hyphal growth at 8 h.p.i. is indicated above each species. Briefly, (-) - no germination, highly-localized hypersensitive response; (+) thru (+++++) > 50% germination. Additionally: (+) hyphae ¼ - ½ length of spore, pinheadsized brown lesions; (++) - hyphae ½ - 1 length of spore, pinhead-sized lesions; (+++) - hyphae 1 -2 spore lengths, larger lesions; (++++) -hyphae 2 -3 spore lengths, lesions coalescing, maceration evident; (+++++) - hyphae > 3 spore lengths, lesions coalescing, maceration evident. For a more complete description of scoring see [Tewari et al., 2003].

not be valid to pool 8 and 48 hpi data from the timecourses in Figure 4.,with data from Figure 6, because in Figure 4 the data were normalized relative to the maximum value of 7 time points in a timecourse for *P. sativum* only, while the data in Figure 6 were normalized relative to the maximum value for all species at 8 and 48 hpi.)

## DISCUSSION
### Expression, as well as sequence, can evolve
To the best of our knowledge, this is the first study to compare the evolution of sequence and expression of multigene family members among closely-related species. We have sequenced orthologous copies of PR10.1, PR10.2 and PR10.3 in four *Pisum* species.

Although our expression data indicate that PR10.4 and PR10.5 orthologues also exist across *Pisum* species, sequences for those orthologues were not obtained from wild pea species. PR10.1, PR10.2 and PR10.3 exhibit almost no amino acid substitutions, when each gene is compared with its orthologue among *Pisum* species. Even intron sequences are highly conserved within orthologues in each species. In contrast, expression patterns among orthologous genes can vary significantly from species to species.

The PR10.1-PR10.3 clade clusters with several genes from *Medicago* species (Figure 2). However, differentiation of this clade into PR10.1/PR10.2 and PR10.3 orthologues appears to be a more recent event, not found in bean or soybean, and possibly unique to *Pisum*. In contrast, both pea PR10.4 and PR10.5 are as closely-related to homologues in other species as to the PR10.1-PR10.3 clade, suggesting that PR10.4 diverged from PR10.5 in some ancestral legume.

It is therefore interesting to note that, in contrast to PR10.1 and PR10.2, PR10.4 and PR10.5 do not show as drastic a degree of divergence in expression among *Pisum* species. The question raised here is: Do some multigene family members become fixed in their expression over time? If so, the more stable a multigene famliy member remains, the more other cellular processes could come to depend upon its expression. In this context, newly-duplicated copies of a gene would be under fewer constraints, and their expression would therefore be more at liberty to evolve.

The elements of gene expression that remain conserved across species may be as informative as those that diverge. PR10.4 and PR10.5 exhibit strong expression in response to *F. solani* in all *Pisum* species. In the same experiments, PR10.3 was not detectable using the RT-PCR assay. This result is consistent with our previous report [Tewari et al. 2003] that PR10.3-specific probe consistently detected lower signal than a PR10.1/PR10.2 subfamily probe, in RNA from *Fusarium solani*-inoculated pea tissue. We therefore attribute the lack of PR10.3 signal in Figure 4 -6 to lower sensitivity of the RT-PCR assay. We can not rule out the possibility that some of the hybridization seen in that paper using a PR10.3-specific probe represented cross-hybridization with PR10.1 or PR10.2 RNA, despite the fact that PR10.1 and PR10.2 control DNAs on the same filters were not detected. If cross hybridization did occur, then those experiments overestimated the amount of PR10.3 transcript present in *Fusarium*-treated pod tissue.

Mylona *et al.*, [1994] have independently cloned the pea PR10.3 cDNA while isolating genes expressed in root epidermis and root-hairs. PR10.3 (referred to as RH2 in reference cited) transcript was far more abundant in roots than transcripts detected using PR10.1-specific oligonucleotides. Further, inoculation of roots with *Rhizobium leguminosarum* bv. *viciae* did not have any detectable effect on the already high PR10.3 transcript accumulation, but caused a slight increase in accumulation of PR10.1 transcript over control levels. Recently, Savouré et al. [1997] demonstrated that PR10 genes in the legume *Medicago sativa* are induced by Nod (nodulation) factors in suspension culture, but expressed constitutively in roots. In contrast, Gamas et al. [1996] have identified PR10 genes in *Medicago truncatula* that are induced during nodule development, but not expressed in roots. While the latter two studies did not use gene-specific probes, they do provide further evidence that gene expression patterns for PR10 genes change from species to species, both with respect to development and to plant/microbe interactions.

Figure 2 suggests that *P. elatius* has at least two copies of PR10.3, which appear to be the result of a recent duplication. Low PR10.3 expression in this species must therefore be conserved for both copies of this gene.

**Does recent duplication imply recent establishment of expression patterns?**

Since gene expression depends in part on regulatory sequences, one *a priori* expectation would be that the more closely related two family members are, the more similar their expression should be. This would be particularly true if similarity between family members was a result of gene conversion. It is difficult to make a case for either for or against this model, with respect to the PR10 family in *Pisum*. In this model, PR10.1 and PR10.2 should have the most similar expression patterns, since they are the most closely-related pair of genes. Yet, across 4 species, PR10.1 and PR10.2 exhibit similar expression in five host/pathogen combinations, but distinctly different patterns in three others.

Clustering of *Medicago* genes separately from PR10.1,2 &3 (Figure 2) suggests that while this group existed in the common ancestor of *Medicago* and *Pisum*, these genes had not further differentiated in that ancestor. PR10.1 and PR10.2 genes cluster separately for all species, suggesting that duplication occurred in the common ancestor of *Pisum* species. We conclude that the subfamiles defined by PR10.1, PR10.2 and PR10.3 are recent. Yet, PR10.1 and PR10.2 expression patterns are the most obviously divergent. If these gene copies are most recent, then their expression patterns must have been established recently as well.

**Regulatory polymorphism: a source of phenotypic diversity?**

Having detected changes in PR10 expression between *P. sativum* and its closest relative, it is apparent that expression patterns can change very rapidly within a multigene family. It is also possible that the rapid evolution of PR10 family expression in response to pathogens is atypical, as multigene families go. For multigene families of other kinds, such as developmentally regulated genes, perhaps evolution is much less rapid. Viewed another way, it may be that rapid evolution of gene expression is most useful, from an evolutionary perspective, in the context of plant/pathogen coevolution.

We have previously shown that infection phenotype diverges among *Pisum* species, with *P. sativum* allowing almost no germation of *F. solani* f. sp. *phaseoli*, contrasted with *P. fulvum* allowing > 50% germination and hyphal growth to 2-3 times the length of the macroconidiospore within 8 h.p.i. [Tewari et al., 2003]. To a first approximation, PR10.4 and PR10.5 show comparatively little change in pathogen-inducible expression, across species, while PR10.1 and PR10.2 show extensive change. Most notably, PR10.1 and PR10.2 expression is only strong at 8 h.p.i in *P. sativum*, which is also most resistant, while these genes show a shift toward later expression in the more susceptible species.

While the data do not provide any causal link between PR10.1/2 expression and the inhibition of germination and hyphal growth, one possiblility is that PR10.1/PR10.2 are controlled solely by a defense pathway, while PR10.4 and PR10.5 are active both in the defense pathway as well as an ABA-inducible pathway (Figure 5). In this model, either PR10.1/PR10.2 have lost an ABA-inducible regulatory element or PR10.4 and PR10.5 have gained an ABA-specific element. In either case, the process being observed could be evidence that PR10 genes can become reassigned over time to different expression regimes.

Taken by itself, the significance of this observation is minimal. Its importance lies in the fact that PR10 is only one of a battery of genes activated during the defense response. Virtually all of the so-called defense genes are present in multigene families. If other defense multigene families also undergo frequent reassignment of regulatory patterns, the effects on disease resistance could be profound.

Changes in regulatory mechanisms such as resistance genes would probably affect whole classes of defense genes simultaneously. Their effects are therefore more likely to be uniform among defense genes, in the manner of an on/off switch. Consequently, variations in disease resistance genes are more likely to result in changes in pathogen specificity, rather than in changes in the type of defense response mounted.

In contrast, mutations in the cis-acting sequences in particular members of a defense multigene family would act on a gene by gene basis. We propose the term "regulatory polymorphism" to refer to changes in expression between alleles of a gene in a population, due to mutations in cis-acting sequences. Regulatory polymorphism at one defense locus by itself might have little effect on the variation of infection phenotype within a plant population. However, the combined effects of polymorphism among many family members of a number of defense multigene families has the potential to generate great phenotypic diversity in a population. A diversity of infection phenotypes in a plant population not only has the potential to slow the spread of a pathogen, but can also provide a genetic basis for natural selection.

The prevalence of regulatory polymorphism, as well as its impact on the evolution of plant/pathogen interactions, remains to be seen. This study has only focused on one gene family. Regulatory polymorphism within other multigene families must be examined to shed further light in the importance of this phenomenon. It is our opinion that the regulatory polymorphism seen in this work represents the tip of an iceberg whose importance, up to now, has been underestimated.

## ACKNOWLEDGEMENTS

This work was supported by Research Grant OGP0105628 from the Natural Sciences and Engineering Research Council of Canada. Support was also provided by the Bank of Nova Scotia. ST was supported as a University of Manitoba Graduate Fellow.